# Modelo híbrido de método *kernel* para computadores quânticos

**Jhordan Silveira de Borba**
sbjhordan@gmail.com
orcid.org/0000-0002-1778-6340
Universidade Federal do Rio Grande do Sul (UFRGS), Porto Alegre, Rio Grande do Sul, Brasil.

**Jonas Maziero**
jonasmaziero@gmail.com
orcid.org/0000-0002-2872-986X
Universidade Federal de Santa Maria (UFSM), Santa Maria, Rio Grande do Sul, Brasil

**RESUMO**

Reconhecendo que a área de aprendizado de máquina quântica é um caminho promissor para oferecer uma revolução nos métodos inteligentes de processamento de dados, propõe-se um método de aprendizado híbrido baseado nos métodos de *kernel* clássicos. Esta proposta também exige que um algoritmo quântico seja desenvolvido para o cálculo de produto interno entre vetores sobre valores contínuos. Para isso ser possível, foi preciso realizar adaptações no método *kernel* clássico, visto que é necessário considerar as limitações impostas pelo espaço de Hilbert do processador quântico. Como um caso de teste, foi verificada a capacidade do algoritmo de aprender a classificar se novos pontos gerados aleatoriamente, em um quadrado finito localizado sob um plano, se encontravam dentro ou fora de um círculo localizado no interior deste quadrado. Verificou-se que o algoritmo foi capaz de detectar corretamente novos pontos em 99% dos casos testados, com uma pequena diferença devido a considerar o raio levemente maior do que o idealizado. O método *kernel* se mostrou capaz de realizar classificações corretamente, assim como o algoritmo do produto interno efetuou satisfatoriamente os cálculos de produto interno utilizando recursos quânticos. Assim, o presente trabalho representa uma contribuição para a área propondo um novo modelo de aprendizado de máquina acessível tanto a físicos quanto a cientistas da computação.

**PALAVRAS-CHAVE**: computação quântica; aprendizado de máquina; classificação binária.

**INTRODUÇÃO**

Quando se fala em métodos de aprendizado de máquina, alguns dos modelos mais populares são redes neurais artificiais. Redes neurais foram responsáveis por avanços significativos e sem precedentes, gerando um enorme impacto no desenvolvimento da área e fazendo jus à sua popularidade (TACCHINO *et al.*, 2019). Outros métodos menos populares são os métodos de *kernel*. Porém, ainda que menos populares, os métodos de *kernel* foram e ainda são extensivamente estudados, tanto teoricamente quanto empiricamente, devido ao seu alto poder de modelagem de dados não lineares. Uma das razões comumente atribuídas à sua menor popularidade é a crença de que seria difícil para os métodos de *kernel* acompanharem as redes neurais artificiais em problemas de larga escala, pois muitos algoritmos *kernel* escalam quadraticamente conforme o número de amostras de treinamento (LU *et al.*, 2014). Entretanto, este é um problema que pode ser superado com o emprego da computação quântica uma vez que o *framework* matemático dos métodos de *kernel* e da computação quântica são bastante parecidos, esta similaridade matemática é útil para o aprendizado de máquinas quânticas. Segundo Schuld (2022), ambos são baseados sob o mesmo princípio. Fica assim evidente a importância que os métodos de *kernel* possuem para a área. Como exemplo de trabalho recente neste sentido, podemos citar o artigo de Chatterjee (2017) em que foi demonstrado como o uso de máquinas de vetores de suporte baseados em estados coerentes reduziu substancialmente o tempo computacional se comparado ao seu algoritmo clássico equivalente. Seguindo esta linha de investigação, o presente trabalho propôs a adição de espaços de estados quânticos, que aumentam exponencialmente a capacidade de representação dos dados.

Aplicações práticas atuais dos métodos de aprendizado de máquina são executadas usualmente como algoritmos clássicos em computadores convencionais (TACCHINO *et al.*, 2019), mas há um crescente aumento de interesse em sua implementação em dispositivos quânticos. Dado que a computação quântica já é uma tecnologia emergente, dentro desse cenário, pesquisas na área de aprendizado de máquina quântica apresentam ser um caminho promissor para futuramente oferecer uma revolução nos métodos inteligentes de processamento de dados (SCHULD; SINAYSIY; PETRUCCIONE, 2015).

Sendo os modelos híbridos quântico-clássico o estado da arte da área (WINCI *et al.*, 2020), o objetivo deste trabalho é propor um modelo híbrido de aprendizado de máquina baseado nos métodos de *kernel*. Dentro desse escopo, devido a sua natureza híbrida, foi necessário propor um algoritmo quântico para o cálculo de produto interno. Buscou-se manter ambas as propostas tão simples quanto possível, visando ser um método de interesse tanto para físicos quanto para cientistas da computação, sendo assim, o presente trabalho pode servir como um primeiro contato para aqueles interessados na área.

**METODOLOGIA**

O método *kernel* clássico tem como principal objetivo capturar padrões não lineares nos dados mapeando-os em dimensões mais elevadas, no qual exibem um padrão linear (CARPUAT, 2017). Isto é feito à custa de um aumento no custo computacional necessário para realizar o treinamento. Porém, uma das vantagens que a computação quântica nos oferece em comparação à computação clássica é o aumento no poder computacional, uma vez que a quantidade de coeficientes do estado quântico aumenta exponencialmente com a quantidade de qubits (bits quânticos), e a natureza da mecânica quântica nos permite utilizar recursos de paralelismo.

O problema que o método *kernel* se propõe a solucionar pode ser enunciado da seguinte forma: tendo duas classes de objetos, para um novo objeto é necessário classificá-lo em uma das classes (SCHÖLKOPF *et al.*, 2002). Considerando então os seguintes dados empíricos:

$$(x_1, y_1)\ldots(x_m, y_m) \in \chi \times \{\pm 1\}, \tag{1}$$

onde $\chi$ é o conjunto onde os padrões $x_i$ (entradas) são retiradas e $\{\pm 1\}$ é o conjunto de onde as saídas $y_i$ são retiradas. Pode-se notar que há somente duas classes, ou seja, um padrão de reconhecimento ou classificação binária. É necessário um tipo adicional de estrutura, para generalizar os dados que não estamos vendo. De maneira esquematizada: Dado um novo padrão de entrada de treinamento $x \in \chi$ queremos predizer a correspondente saída $y \in \{\pm 1\}$. Fazendo isso para cada $x \in \chi$, nos leva a estimar uma função $f: \chi \to \{\pm 1\}$. Então, para um novo padrão de entrada, queremos escolher a saída que seja coerente com os exemplos de treinamento.

Por isso, é necessário usar noções de similaridade do conjunto de dados de entrada e do conjunto de dados de saída. A similaridade da saída é obtida de forma bastante direta. Como há somente duas classes, os rótulos podem ser apenas idênticos entre si ou diferentes. Já a similaridade das entradas é mais complicada de se obter e constitui o núcleo do aprendizado de máquina.

Consideramos uma medida de similaridade na forma:

$$\begin{aligned} k: \chi \times \chi &\to \Re \\ (x_i, x_j) &\to \left(k(x_i, x_j)\right) \end{aligned} \tag{2}$$

Dados dois padrões de entrada, a função $k$ retorna um número real, caracterizando sua similaridade. A função $k$ é assumida ser simétrica, $\left(k(x_i, x_j) = k(x_j, x_i)\right)$, e é denominada como *kernel*. Medidas gerais de similaridade são difíceis de serem estudadas. Um caso particular e simples de similaridade é o produto interno. Neste trabalho, utilizamos uma medida baseada no produto interno, como será discutido adiante. Então, é necessário representar as entradas como um vetor em algum espaço que possua produto interno. Para isso, usa-se o mapa:

$$\begin{aligned} \Phi = \chi &\to H \\ x &\to |x\rangle = \Phi(x) \end{aligned} \tag{3}$$

O espaço com produto interno $H$ é tipicamente chamado de espaço de características. Mesmo que os padrões originais existam em um espaço que há produto interno, pode ser desejável usar medidas de similaridade mais gerais, obtidas com a aplicação de um mapa. Nesse caso, normalmente será utilizado um mapa não linear. Um algoritmo característico de reconhecimento de padrões do método *kernel* parte do cálculo da média das duas classes:

$$|c_+\rangle = \frac{1}{m_+}\sum_{\{i|y_i=+1\}}|x_i\rangle$$
$$|c_-\rangle = \frac{1}{m_-}\sum_{\{i|y_i=-1\}}|x_i\rangle \quad (4)$$

onde $m_\pm$ é o número de entradas com a respectiva saída $\pm 1$. Basicamente, o que o algoritmo faz é associar uma nova entrada $|x\rangle$ à classe com a média mais próxima. Para uma função *kernel* genérica, $k(x_i, x_j)$, a classe que a entrada deve ser classificada vai depender do sinal da seguinte equação (SCHÖLKOPF *et al.*, 2002):

$$y = \text{sign}\left[\frac{1}{m_+}\sum_{\{i|y_i=+1\}}k(x, x_i) - \frac{1}{m_-}\sum_{\{i|y_i=-1\}}k(x, x_i) + b\right]$$
$$b = \frac{1}{2}\left[\frac{1}{m_-^2}\sum_{\{i,j|y_i=y_j=-1\}}k(x_i, x_j) - \frac{1}{m_+^2}\sum_{\{i,j|y_i=y_j=+1\}}k(x_i, x_j)\right] \quad (5)$$

onde $b$ é chamado de deslocamento.

O mapa $\Phi$ depende do problema que está sendo solucionado. Por isso, será necessário propor um problema para avançar com o presente trabalho. Interessados no método de *kernel* clássico podem consultar (HERBRICH, 2001) (HOFMANN; SCHÖLKOPF; SMOLA, 2008) (HOFMANN; SCHÖLKOPF; SMOLA, 2006) (SCHÖLKOPF et al, 2002). O problema escolhido como exemplo será a classificação de novos pontos gerados aleatoriamente dentro de uma região quadrada em duas dimensões. O método deve ser capaz de classificar se estes novos pontos estão no interior ou no exterior de um círculo contido e centralizado neste quadrado. Em duas dimensões esse é um problema não linear. Em dimensões superiores esse problema pode ser tratado linearmente. Será utilizado o espaço do processador quântico como espaço de dimensão superior. Utilizando um sistema com dois bits quânticos (*qubits*), o mapa deve ser responsável por levar os dados de entrada do sistema em duas dimensões, para o espaço com quatro dimensões do processador quântico. Para isto, será utilizada a equação geral do cone:

$$\left(\frac{x}{a}\right)^2 + \left(\frac{y}{b}\right)^2 = \left(\frac{z}{c}\right)^2. \quad (6)$$

Para projetar pontos do plano original em duas dimensões sob a superfície do cone no espaço de 3 dimensões, pode-se utilizar:

$$f(x,y) = z = c\sqrt{\left(\frac{x}{a}\right)^2 + \left(\frac{y}{b}\right)^2}. \quad (7)$$

A **Figura 1** ilustra o processo de transformação do problema não linear em um problema linear em três dimensões através do mapa, lembrando ainda existe uma quarta dimensão no processador quântico que é utilizada para questões de normalização.

Figura 1 - Ilustração do problema proposto

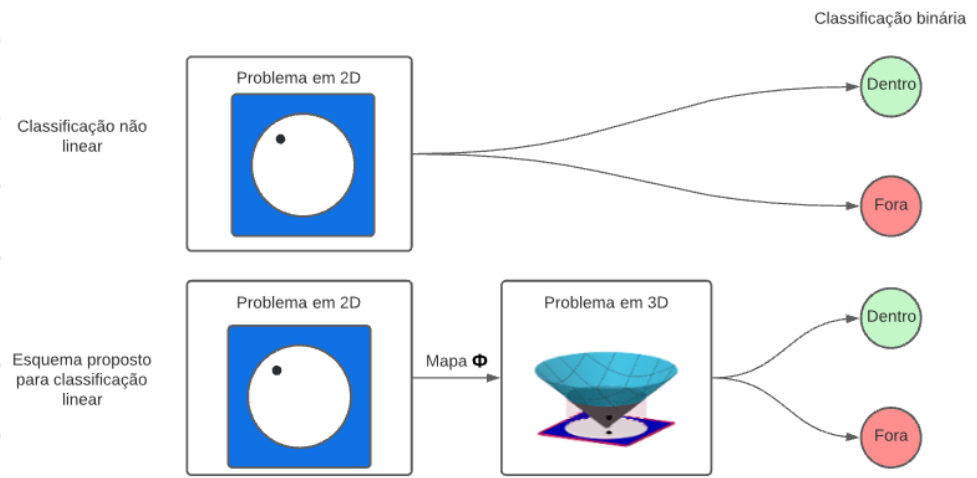

Fonte: Autoria própria.

Neste ponto é necessário prestar atenção em algumas condições que a Mecânica Quântica impõe, dado que o problema será resolvido no espaço do processador quântico em quatro dimensões. Um estado qualquer do sistema de dois *qubits* é dado por:

$$|\psi\rangle = \sum_{i=0}^{3} a_i |i\rangle, \quad (8)$$

onde os termos $a_i$ são chamados de amplitudes de probabilidade ou coeficientes. Pela condição de normalização quando considerado como um número real, tem-se que $\sum_{i=0}^{3} a_i^2 = 1$, o que implica que cada coeficiente é necessariamente menor ou igual a 1. Utilizando o quarto coeficiente para satisfazer as condições de normalização, os dados de entrada serão levados para o espaço de 4 dimensões formado pelos 2 *qubits* da seguinte forma:

$$\begin{aligned} a_0 &= x, \\ a_1 &= y, \\ a_2 &= \sqrt{\frac{x^2+y^2}{2}}, \\ a_3 &= \sqrt{(1 - a_0^2 - a_1^2 - a_2^2)}, \end{aligned} \quad (9)$$

onde foram escolhidos os parâmetros $(a, b, c) = \left(\sqrt{(2)}, \sqrt{(2)}, 1\right)$. Esta escolha deve-se ao fato de que o valor máximo que três parâmetros podem assumir simultaneamente devido à condição de normalização é $1/\sqrt{(3)}$, e com este conjunto de parâmetros, para $(a_0, a_1) = \left(\pm 1/\sqrt{(3)}, \pm 1/\sqrt{(3)}\right)$, temos que $a_2 = 1/\sqrt{(3)}$. Por fim, a função *kernel* utilizada será o valor medido durante a medição no computador quântico, isto é $k = \left|\langle x_i | x_j \rangle\right|^2$.

Para o cálculo do produto interno no computador quântico, dados dois vetores de dimensão $m$, estes devem ser codificados usando $m$ coeficientes para definir uma função de onda geral. Na prática, dados dois vetores arbitrários:

$$\vec{x_i} = \begin{pmatrix} i_0 \\ i_1 \\ \dots \\ i_{m-1} \end{pmatrix}, \vec{x_w} = \begin{pmatrix} w_0 \\ w_1 \\ \dots \\ w_{m-1} \end{pmatrix}, \qquad (10)$$

podemos obter o quadrado do produto interno entre ambos:

$$|\vec{x_i} \cdot \vec{x_w}|^2 = m^2 \langle \psi_i | \psi_w \rangle^2. \qquad (11)$$

Para isto, aplicam-se dois operadores unitários sob o sistema. Estes operadores devem possuir as seguintes propriedades $U_i|00\dots00\rangle = |\psi_i\rangle$ e $U_w|\psi_w\rangle = |11\dots11\rangle$, onde o estado $|\psi_j\rangle$ é o vetor $\vec{x_j}$ codificado. Isto é:

$$|\psi_j\rangle = \frac{1}{\sqrt{m}} \sum_{k=0}^{m-1} j_k |k\rangle, \qquad (12)$$

sendo $k$ o decimal formado a partir da respectiva *string* binária. Por exemplo, $0 = 00\dots00, 1 = 00\dots01, etc$. Visto que os operadores satisfazem estas propriedades, então (TACCHINO *et al.*, 2019):

$$\langle \psi_w | \psi_i \rangle = c_{m-1}, \qquad (13)$$

onde $c_{m-1}$ é o coeficiente de índice $m-1$ do estado do sistema após a aplicação de ambos operadores:

$$U_w U_i |0\rangle = \sum_{j=0}^{m-1} c_j |j\rangle. \qquad (14)$$

Antes de avançar com a discussão, vale ressaltar que após a aplicação de ambos operadores, pode-se utilizar um qubit auxiliar $|a\rangle$ e um operador NOT multi-controlado entre os N qubits para codificar a informação no qubit alvo:

$$U_w U_i |0\rangle |0\rangle_a \to \sum_{j=0}^{m-2} c_j |j\rangle |0\rangle_a + c_{m-1} |m-1\rangle |1\rangle_a. \qquad (15)$$

Dessa forma é necessário apenas medir a probabilidade do *qubit* auxiliar estar no estado 1.

A próxima etapa é construir os operadores unitários a fim de que exibam as propriedades desejadas. É necessário apenas que o operador $U_i$ seja representado por uma matriz unitária $m \times m$ com o vetor $\vec{x_i}$ na sua primeira coluna. E de forma análoga, $U_w$ deve possuir o conjugado do vetor $\vec{x_j}$ em sua última linha, ou, como vamos trabalhar apenas com valores reais, o próprio vetor $\vec{x_j}$ em sua última linha.

A proposta do algoritmo responsável pela implementação dos operadores unitários e, consequentemente, o cálculo do produto interno no computador quântico, é feita a partir de "blocos" de variação do módulo do coeficiente. Isto é, cada bloco deve ser responsável por alterar o módulo de um coeficiente em específico, mantendo os outros inalterados. Devido à exigência de normalização, tanto do sistema como de cada *qubit* que compõe o sistema, as variações de módulo dos coeficientes ocorrem sempre em pares. Por isso, é separado um coeficiente para servir apenas de controle e garantir a normalização.

Escolhendo o último coeficiente como controle, a seguinte sequência de operações $M_0 = (X \otimes I) \cdot cX_0 \cdot cX_1 \cdot cu_3 \cdot cX_1 \cdot (X \otimes I)$, modifica o par de coeficientes $(c_0, c_3)$, lembrando que para dois *qubits* temos 4 coeficientes.

Portanto, $c_3$ é o *qubit* de controle. Além disso, vale destacar que $X$ é um operador NOT, $cu_3$ é a versão controlada do operador $u_3(\theta, \varphi, \lambda)$ (rotação de um único *qubit* conforme 3 ângulos de Euler), com os ângulos omitidos por conveniência. E por último, $cX_i$ é um $X$ controlado utilizando o *qubit* $i$ como controle.

Para o segundo *qubit*, pode-se utilizar apenas $u_3$. Para o terceiro *qubit*, tem-se $M_2 = cX_1 \cdot cX_0 \cdot cX_1 \cdot cu_3 \cdot cX_1 \cdot cX_0 \cdot cX_1$. Assim, denotando os ângulos por $\{a, b, c\}$, os três blocos possuem a seguinte representação:

$$M_0 = \begin{pmatrix} \cos\left(\frac{a}{2}\right) & 0 & 0 & -\sin\left(\frac{a}{2}\right) \\ 0 & 1 & 0 & 0 \\ 0 & 0 & 1 & 0 \\ \sin\left(\frac{a}{2}\right) & 0 & 0 & \cos\left(\frac{a}{2}\right) \end{pmatrix}, \quad (16)$$

$$M_1 = \begin{pmatrix} 1 & 0 & 0 & 0 \\ 0 & \cos\left(\frac{b}{2}\right) & 0 & -\sin\left(\frac{b}{2}\right) \\ 0 & 0 & 1 & 0 \\ 0 & \sin\left(\frac{b}{2}\right) & 0 & \cos\left(\frac{b}{2}\right) \end{pmatrix}, \quad (17)$$

$$M_2 = \begin{pmatrix} 1 & 0 & 0 & 0 \\ 0 & 1 & 0 & 0 \\ 0 & 0 & \cos\left(\frac{c}{2}\right) & -\sin\left(\frac{c}{2}\right) \\ 0 & 0 & \sin\left(\frac{c}{2}\right) & \cos\left(\frac{c}{2}\right) \end{pmatrix}. \quad (18)$$

Cada uma dessas matrizes altera a amplitude de apenas um dos coeficientes, além do coeficiente de controle. Quando aplicamos os três blocos na seguinte sequência $U_a = M_2 \cdot M_1 \cdot M_0$, em um estado inicial $|\psi_0\rangle = [1 \ \ 0 \ \ 0 \ \ 0]^T$, temos $|a_0\rangle = U_a|\psi_0\rangle$. Segundo a propriedade desejada para o operador, $U_i|00\ldots00\rangle = |\psi_i\rangle$, onde

$$|a\rangle = \begin{pmatrix} \cos\left(\frac{a}{2}\right) \\ -\sin\left(\frac{a}{2}\right)\sin\left(\frac{b}{2}\right) \\ -\sin\left(\frac{a}{2}\right)\cos\left(\frac{b}{2}\right)\sin\left(\frac{c}{2}\right) \\ \sin\left(\frac{a}{2}\right)\cos\left(\frac{b}{2}\right)\cos\left(\frac{c}{2}\right) \end{pmatrix} = \begin{pmatrix} a_0 \\ a_1 \\ a_2 \\ a_3 \end{pmatrix}. \quad (19)$$

Logo, sabendo os coeficientes $a_i$ do vetor $|a\rangle$, é possível obter os ângulos através das seguintes relações:

$$\begin{aligned} a &= 2\arccos(a_0), \\ b &= 2\arcsin\left(-\frac{a_1}{\sin(a/2)}\right), \\ c &= 2\arccos\left(-\frac{a_2}{\sin(a/2)\cos(b/2)}\right). \end{aligned} \quad (20)$$

Agora aplicando o mesmo operador $U_a = U_b$, porém em um estado $|11\ldots11\rangle$, vem que $U_b|11\ldots11\rangle = |b\rangle$, sendo:

$$|b\rangle = \begin{bmatrix} \sin\left(\frac{a}{2}\right)\cos\left(\frac{b}{2}\right)\cos\left(\frac{c}{2}\right) \\ \sin\left(\frac{b}{2}\right)\cos\left(\frac{c}{2}\right) \\ \sin\left(\frac{c}{2}\right) \\ \cos\left(\frac{a}{2}\right)\cos\left(\frac{b}{2}\right)\cos\left(\frac{c}{2}\right) \end{bmatrix} = \begin{bmatrix} b_0 \\ b_1 \\ b_2 \\ b_3 \end{bmatrix}. \quad (21)$$

Sendo $\overline{U}_w^T = U_b$, temos a propriedade do segundo operador satisfeita: $U_w|\psi_w\rangle = |11\ldots11\rangle$. Então, de maneira análoga a $U_i$, podemos utilizar os blocos $M_j$ para a construção do operador $U_w$, tendo seus componentes codificados da seguinte forma:

$$\begin{aligned} c &= 2\arcsin(b_2), \\ b &= 2\arcsin\left(\frac{b_1}{\cos(c/2)}\right), \\ a &= 2\arcsin\left(\frac{a_0}{\cos(b/2)\cos(c/2)}\right). \end{aligned} \quad (22)$$

Aplicando a sequência de operadores $U_b \cdot U_a$ no estado inicial, uma vez que a primeira coluna de $U_a$ é $|a\rangle$ e a última linha de $U_b$ é $\langle b|$, vem que $|\psi\rangle = U_b \cdot U_a|\psi_0\rangle$. Explicitamente:

$$|\psi\rangle = \begin{pmatrix} A_{11} & A_{12} & A_{13} & A_{14} \\ A_{21} & A_{22} & A_{23} & A_{24} \\ A_{31} & A_{32} & A_{33} & A_{34} \\ b_0 & b_1 & b_2 & b_3 \end{pmatrix} \begin{pmatrix} a_0 & A_{12} & A_{13} & A_{14} \\ a_1 & A_{22} & A_{23} & A_{24} \\ a_2 & A_{32} & A_{33} & A_{34} \\ a_3 & A_{42} & A_{43} & A_{44} \end{pmatrix} \begin{pmatrix} 1 \\ 0 \\ 0 \\ 0 \end{pmatrix}. \quad (23)$$

$$|\psi\rangle = \begin{pmatrix} A_{11} & A_{12} & A_{13} & A_{14} \\ A_{21} & A_{22} & A_{23} & A_{24} \\ A_{31} & A_{32} & A_{33} & A_{34} \\ b_0 & b_1 & b_2 & b_3 \end{pmatrix} \begin{pmatrix} a_0 \\ a_1 \\ a_2 \\ a_3 \end{pmatrix}. \quad (24)$$

Então:

$$|\psi\rangle = [c_0 \quad c_1 \quad c_2 \quad \sum_i^3 a_i b_i]^T. \quad (25)$$

Logo, a última componente do estado final do sistema é $c_3 = \langle b|a\rangle$, como desejado. Consequentemente o valor medido será $|c_3|^2 = |\langle a|b\rangle|^2$. Antes de encerrar esta discussão, é necessário prestar atenção a alguns detalhes. A ordem de aplicação dos operadores $M_i$ não é arbitrária, e o conjunto de equações utilizadas para calcular os ângulos $\theta_i$ no bloco $M_i$ pode apresentar problemas sob determinadas circunstâncias. Por exemplo, no conjunto de equações para o operador $U_a$, tem-se que $-sin\left(\frac{a}{2}\right)sin\left(\frac{b}{2}\right) = a_1$ mas se $\left|sin\left(\frac{a}{2}\right)\right| < a_1$ então $\left|\frac{a_1}{\sin(a/2)}\right| > 1$, o que leva à conclusão de que $sin\left(\frac{b}{2}\right) > 1$. Para evitar essa categoria de problemas, diferentes sequências de operações devem ser aplicadas em diferentes situações. Para o vetor $|a\rangle$, no caso geral em que $|a_i| \leq |a_j| \leq |a_k|$ deve-se aplicar $M_k \cdot M_j \cdot M_i \cdot (X \otimes X)$ na construção de $U_a$. O vetor $|b\rangle$, tendo a mesma sequência genérica $|b_i| \leq |b_j| \leq |b_k|$, utiliza-se a ordem inversa, isto é: $M_i \cdot M_j \cdot M_k$ para a construção de $U_b$. Observamos que os ângulos $a, b, c$ devem ser também recalculados caso a caso.

**RESULTADOS**

Todos os algoritmos foram implementados em Python utilizando a biblioteca NumPy para cálculos analíticos e o *kit* de desenvolvimento Qiskit para Python. Qiskit tem uma importância especial para a área, uma vez que é um *kit* de desenvolvimento de código aberto para trabalhar com computadores quânticos, tanto em simuladores quanto em dispositivos reais da IBM, e que atualmente possui bons tutoriais de acesso livre. Podemos citar como exemplo a sequência escrita por Koch, Wessing, Alsing (2019) e Koch *et al.* (2020).

Os códigos foram executados a partir do recurso de computação em nuvem disponibilizado pelo Google, chamado Google Colab. Através desta plataforma, o Google disponibiliza um processador Intel(R) Xeon(R) CPU @ 2.30GHz com 2 núcleos e 12GB de memória RAM. Neste ambiente, também foi utilizada a versão 0.16.1 do Qiskit, que possui as versões de seus módulos listadas no **Quadro 1**.

Quadro 1 – Módulos utilizados

| Módulo | Versão |
|---|---|
| qiskit-aer | 0.7.2 |
| qiskit-aqua | 0.8.1 |
| qiskit-ibmq-provider | 0.11.1 |
| qiskit-ignis | 0.5.1 |
| qiskit-terra | 0.16.1 |

Fonte: Autoria própria.

Por fim, foi utilizado o simulador disponibilizado pelo Qiskit, chamado '*statevector_simulator*'. Este simulador não inclui ruídos, ele é um simulador de um circuito quântico ideal. Para satisfazer as condições discutidas anteriormente, o quadrado centralizado na origem possui um lado de 1.08, isto é, os pontos podem ser gerados aleatoriamente nas coordenadas $(x, y)$, sendo $x, y \in [-0{,}54, 0{,}54]$.

Para o conjunto de treinamento, foi gerada uma grade de pontos com espaçamento $\Delta x = \Delta y = 0.04$ entre si e o raio do círculo utilizado para calcular as saídas ideais foi 0.42. Os pontos que estão fora do círculo foram considerados compondo a classe positiva (+1) enquanto os pontos no interior compõem a classe negativa (-1). Após a etapa de definição das saídas ideais para cada entrada do conjunto de treinamento, o raio é esquecido.

Utilizando estes critérios, foi calculado um deslocamento $b \approx 0.15$. A partir deste valor de deslocamento para 100 novos pontos aleatórios no interior do quadrado, 99% dos pontos foram classificados corretamente. Mais especificamente, 100% dos pontos no interior e 98% dos pontos no exterior foram classificados corretamente. Isso sugere que o círculo que o método *kernel* aprendeu possui um raio ligeiramente maior do que o utilizado para gerar as saídas ideais do conjunto de dados de treinamento.

**CONCLUSÕES**

O modelo híbrido de aprendizado de máquina implementado no simulador disponibilizado pelo Qiskit apresentou um resultado satisfatório que demonstra um bom potencial quanto à capacidade de aprendizagem. Algumas otimizações que poderiam ser implementadas dizem respeito ao seu aspecto clássico, tais como um algoritmo *kernel* híbrido para situações mais gerais.

Quanto aos objetivos iniciais, o algoritmo quântico proposto para o cálculo de produto interno de estados de dois *qubits* mostrou-se eficiente, sendo capaz de executar satisfatoriamente os cálculos necessários no simulador. E o modelo *kernel* também apresentou bons resultados, classificando satisfatoriamente novas entradas que não constavam nos dados de treinamento.

Em adição a estes resultados, o objetivo foi alcançado com um modelo proposto relativamente simples e acessível a novos interessados que venham tanto da computação quanto da física. Este resultado também merece destaque uma vez que os métodos de *kernel* se apresentam como um método capaz de resolver de forma linear os problemas não-lineares, sendo uma alternativa ao problema dado que a computação quântica se baseia na mecânica quântica, que, no que lhe concerne, é linear. Então o computador quântico também opera linearmente. Portanto, acreditamos que esse trabalho representa mais um passo no caminho certo do desenvolvimento da inteligência artificial quântica.

Os computadores quânticos atuais têm taxas de ruído muito altas para possibilitar a verificação experimental dos circuitos quânticos apresentados neste artigo. Embora a simulação tenha mostrado o funcionamento desses circuitos no caso ideal, futuramente seria interessante também utilizar simuladores quânticos que incluem o ruído na simulação. Isso poderá fornecer uma visão mais clara sobre a implementabilidade experimental desses circuitos num futuro próximo, quando as taxas de ruídos forem diminuídas nos computadores quânticos reais.

# Hybrid model of the kernel method for quantum computers

**ABSTRACT**

The field of quantum machine learning is a promising way to lead to a revolution in intelligent data processing methods. In this way, a hybrid learning method based on classic kernel methods is proposed. This proposal also requires the development of a quantum algorithm for the calculation of internal products between vectors of continuous values. In order for this to be possible, it was necessary to make adaptations to the classic kernel method, since it is necessary to consider the limitations imposed by the Hilbert space of the quantum processor. As a test case, we applied this new algorithm to learn to classify whether new points generated randomly, in a finite square located under a plane, were found inside or outside a circle located inside this square. It was found that the algorithm was able to correctly detect new points in 99% of the samples tested, with a small difference due to considering the radius slightly larger than the ideal. However, the kernel method was able to perform classifications correctly, as well as the internal product algorithm successfully performed the internal product calculations using quantum resources. Thus, the present work represents a contribution to the area, proposing a new model of machine learning accessible to both physicists and computer scientists.

**KEYWORDS:** quantum computing; machine learning; binary classification.

# Modelo híbrido del método del *kernel* para computadoras cuánticas

**RESUMEN**

Reconociendo que el área de aprendizaje de la máquina cuántica es una forma prometedora de ofrecer una revolución en los métodos inteligentes de procesamiento de datos, en este trabajo proponemos un método de aprendizaje híbrido basado en métodos clásicos del *kernel*. Esta propuesta también requiere que se desarrolle un algoritmo cuántico para el cálculo del producto interno entre vectores de valores continuos. Para que esto fuera posible, fue necesario adaptar el método clásico del *kernel*, ya que es necesario considerar las limitaciones impuestas por el espacio de Hilbert del procesador cuántico. Como caso de prueba, fue testada la capacidad del algoritmo para aprender a clasificar si nuevos puntos generados aleatoriamente, en un cuadrado finito ubicado en un plano, se encontraron dentro o fuera de un círculo ubicado dentro de este cuadrado. Se encontró que el algoritmo fue capaz de detectar correctamente nuevos puntos para 99% de las puntos probados, con una pequeña diferencia por considerar el radio un poco más grande que el ideal. Sin embargo, el método del *kernel* fue capaz de realizar clasificaciones correctamente, así como el algoritmo de producto interno realizó satisfactoriamente los cálculos de producto interno utilizando recursos cuánticos. Así, el presente trabajo representa un aporte al área, proponiendo un nuevo modelo de aprendizaje automático accesible tanto a físicos como a informáticos.

**PALABRAS CLAVE:** computación cuántica; aprendizaje de máquina; clasificación binaria.

# REFERÊNCIAS